\documentclass[12pt]{article}
\usepackage{hyperref}
\usepackage[english]{babel}
\usepackage{amsmath}
\usepackage{latexsym}
\usepackage{amssymb}
\usepackage[all]{xy}
\usepackage[dvips]{graphicx}
\usepackage{epsfig}
\usepackage{psfrag}

\numberwithin{equation}{section} \numberwithin{table}{section}
\numberwithin{figure}{section}

\begin{document}


\begin{titlepage}
  \begin{flushright}
  {\small CQUeST-2010-0377}
  \end{flushright}

  \begin{center}

    \vspace{20mm}

    {\LARGE \bf Notes on Properties of Holographic \\Strange Metals}

    \vspace{10mm}

   Bum-Hoon Lee$^{\ast\dag}$ and Da-Wei Pang$^{\dag}$

    \vspace{5mm}
    {\small \sl $\ast$ Department of Physics, Sogang University}\\
    {\small \sl $\dag$ Center for Quantum Spacetime, Sogang University}\\
    {\small \sl Seoul 121-742, Korea\\}
    {\small \tt bhl@sogang.ac.kr, pangdw@sogang.ac.kr}
    \vspace{10mm}

  \end{center}

\begin{abstract}
\baselineskip=18pt We investigate properties of holographic strange
metals in $p+2$-dimensions, generalizing the analysis performed in
arXiv:0912.1061. The bulk spacetime is $p+2$-dimensional Lifshitz
black hole, while the role of charge carriers is played by probe
D-branes. We mainly focus on massless charge carriers, where most of
the results can be obtained analytically. We obtain exact results
for the free energy and calculate the entropy density, the heat
capacity as well as the speed of sound at low temperature. We obtain
the DC conductivity and DC Hall conductivity and find that the DC
conductivity takes a universal form in the large density limit,
while the Hall conductivity is also universal in all dimensions. We
also study the resistivity in different limits and clarify the
condition for the linear dependence on the temperature, which is a
key feature of strange metals. We show that our results for the DC
conductivity are consistent with those obtained via Kubo formula and
we obtain the charge diffusion constant analytically. The
corresponding properties of massive charge carriers are also
discussed in brief.

\end{abstract}
\setcounter{page}{0}
\end{titlepage}

\pagestyle{plain} \baselineskip=19pt

\tableofcontents

\section{Introduction}
One of the most interesting subjects in condensed matter physics is
to understand the properties of strongly interacting systems of
fermions, in particular, to understand the thermodynamic and
transport properties of the ``strange metal'' phases of heavy
fermion compounds~\cite{Stewart:2001zz} and high temperature
superconductors~\cite{Cooper:2009ra, Hussey:2008ne}. The ``strange
metal'' phases possess various interesting ``non-Fermi liquid''
features. For example, the DC resistivity is linear in temperature
$T$ when $T$ is much less than the chemical potential
$\mu$~\cite{Martin:1990ss}. On the other hand, the AC conductivity
exhibits a non-trivial scaling behavior
$\sigma(\omega)\sim\omega^{-\nu}$ with
$\nu\neq1$~\cite{vandeMarel:2003wn}. Furthermore, the Hall
conductivity also becomes anomalous~\cite{Tyler:1997aw}. It is
widely believed that to gain a better understanding of such
properties requires us to go beyond the regime of weak coupling.

Recently, inspired by the AdS/CFT
correspondence~\cite{Maldacena:1997re, Aharony:1999ti},
investigations on the applications of AdS/CFT to condensed matter
physics (also named as the AdS/CMT correspondence) have been
accelerated enormously~\cite{Hartnoll:2009sz}. Since the AdS/CFT
correspondence provides us with useful tools for understanding the
dynamics of strongly coupled field theories in the dual weakly
coupled gravity side, it may open a new window for studying
real-world physics in the context of holography. Thus one can expect
that the peculiar properties of ``strange metals'' can be well
understood via the AdS/CFT correspondence.

One class of such holographic models of strange metals were proposed
in~\cite{Lee:2008xf, Liu:2009dm, Cubrovic:2009ye, Faulkner:2009wj,
Faulkner:2010da}. In such models the dual bulk spacetime was
described by an extremal RN-${\rm AdS}_{4}$ black hole, whose near
horizon geometry contains an ${\rm AdS}_{2}$ part. It was shown that
the low energy behavior of these non-Fermi liquids was controlled by
the near horizon IR fixed point. Examples whose single-particle
spectral function and transport behavior resembling those of strange
metals were found within this class of models.

Recently a complementary approach was proposed
in~\cite{Hartnoll:2009ns}, where they considered a bulk
gravitational background which was dual to a neutral
Lifshitz-invariant quantum critical theory, while the gapped probe
charge carriers were described by D-branes. The non-Fermi liquid
scalings, such as linear resistivity, observed in strange metal
regimes can be realized by choosing the dynamical critical exponent
$z$ appropriately. They also outlined three distinct string theory
realizations of Lifshitz geometries. Similar investigations on
charged dilaton black holes were carried out
in~\cite{Charmousis:2010zz, Lee:2010ii}, where strange metallic
behavior was found by properly fixing the parameters of the bulk
gravity theory.

In this paper we will study properties of holographic strange metals
by employing the approach proposed in~\cite{Hartnoll:2009ns}. The
bulk theory is chosen to be a $p+2$-dimensional asymptotically
Lifshitz black hole, while the charge carriers are still represented
by probe D$q$-branes. For a practical application in condensed
matter physics, the value of $p$ is fixed. For example, one should
set $p=2$ to study the $(2+1)$-dimensional layered systems. However,
our physical spacetime may be higher dimensional with extra
dimensions, and more spacetime dimensions might be holographically
generated when extra adjoint fields are involved. Therefore, it is
of interest to consider generalizations to various dimensions and
try to find the universal behavior.

We shall consider both massless and massive charge carriers. In the
language of D-brane physics, the massive case is related to a
non-trivial embedding profile of the probe D-brane into bulk
spacetime. We mainly focus on the calculations of massless charge
carriers and we will give some remarks on the massive case. We find
that for massless charge carriers, the chemical potential $\mu$ and
the free energy $\Omega$ can be obtained analytically. Therefore one
can work out other thermodynamic quantities by standard procedures.
It turns out that at low temperature, the specific heat behaves like
$c_{v}\propto T^{2p/z}/{d}$, where $z$ is the dynamical component
and $d$ is related to the charge density. It can resemble a gas of
free bosons, fermions or new types of holographic quantum liquids by
fixing the values of $z$ and $p$. The speed of sound at low
temperature is given by $z/p$. We calculate the DC conductivity and
DC Hall conductivity by introducing corresponding $U(1)$ gauge
fields on the worldvolume of the probe D-branes. In the large charge
density limit, the expression for the DC conductivity is the same as
the four-dimensional counterpart, while it becomes $p$-dependent in
the vanishing charge density limit. On the contrary, the DC Hall
conductivity exhibits universal behavior in general
$p+2$-dimensions. Therefore the linear dependence of temperature $T$
for the resistivity, which is a common feature of strange metals,
can be realized by adjusting parameters in the theory. We re-obtain
the DC conductivity by using the result derived from Kubo formula,
which proves consistency of the two approaches. We also obtain
analytic results for the charge diffusion constant of massless
charge carriers by using a generalized version of the formula
derived from the membrane paradigm. The properties of massive charge
carriers are also discussed in brief.

The rest of the paper is organized as follows. The general setup
will be illustrated in section 2 and the thermodynamics of massless
charge carriers will be discussed in section 3. In section 4 we will
calculate the DC conductivity and Hall conductivity in general
$p+2$-dimensions and discuss various limits of the resistivity . We
will re-obtain the DC conductivity by Kubo formula and the charge
diffusion constant in section 5. Remarks on massive charge carriers
are given in section 6. Finally, summary and discussion are
presented in section 7.

\section{The Setup}
Consider the following ten-dimensional bulk spacetime,
\begin{equation}
ds^{2}=L^{2}[-r^{2z}f(r)dt^{2}+\frac{dr^{2}}{r^{2}f(r)}+r^{2}d{\vec{x}_{p}}^{2}]
+L^{2}d\Omega^{2}_{8-p},
\end{equation}
where the metric of $S^{8-p}$ is given by
\begin{equation}
d\Omega^{2}_{8-p}=d\theta^{2}+\cos^{2}\theta
d\Omega^{2}_{n}+\sin^{2}\theta d\Omega^{2}_{7-p-n}.
\end{equation}
The non-spherical part of this metric is a $p+2$-dimensional
asymptotically Lifshitz black hole with $p\geq2$, whose temperature
is given by
\begin{equation}
T=\frac{r^{z+1}_{+}}{4\pi}\big|f^{\prime}(r_{+})\big|\equiv
\frac{\lambda r^{z}_{+}}{4\pi z},
\end{equation}
where $r_{+}$ denotes the radial position of the horizon. The
parameter $\lambda$ is determined by the explicit form of the
function $f(r)$.

When $f(r)=1$, this part of the metric possesses the following
anisotropic scaling symmetry,
\begin{equation}
t~\rightarrow~\lambda^{z}t,~~~\vec{x}~\rightarrow~\lambda\vec{x},~~~
r~\rightarrow~\frac{r}{\lambda}.
\end{equation}
Such backgrounds can be considered as the gravity duals of Lifshitz
fixed points, which was initially constructed
in~\cite{Kachru:2008yh}. Asymptotically Lifshitz black hole
solutions were investigated in~\cite{Taylor:2008tg}
and~\cite{others}. The relations between Lifshitz black holes and
heavy fermion metals were discussed in~\cite{Brynjolfsson:2010rx}.
It should be emphasized that some of the analytic solutions involve
a non-trivial dilaton even at zero temperature, which breaks the
scaling symmetry of Lifshitz spacetime. Lifshitz spacetimes in
string theory with unconventional scaling symmetry were extensively
studied in~\cite{Azeyanagi: 2009lt}. Since three distinct approaches
for realizing Lifshitz geometries in string theory were proposed
in~\cite{Hartnoll:2009ns}, we may make the above mentioned ansatz in
ten-dimensional spacetime and consider D-brane probes in such
backgrounds.

We assume that the probe D$q$-brane extends all the spacial
directions and wraps the $S^{n}$ part inside $S^{8-p}$, which leads
to $q=p+n+1$. The dynamics of the D$q$-brane is described by the
following DBI action (neglecting the Wess-Zumino term)
\begin{equation}
S_{DBI}=-T_{q}\int dtdrd^{p}xd\Omega_{n}e^{-\phi}\sqrt{-{\rm
det}(g_{ab}+2\pi\alpha^{\prime}F_{ab})},
\end{equation}
where $T_{q}$ denotes the tension of the D$q$-brane,
$T_{q}=1/((2\pi)^{p}g_{s}\ell_{s}^{p+1})$. $g_{ab}$ and $F_{ab}$ are
the induced metric and the $U(1)$ gauge field strength on the
worldvolume of the D$q$-brane. The scaling symmetry of Lifshitz
geometry requires that the dilaton should be constant,
$\phi=\phi_{0}$, thus we can arrive at the effective action
\begin{equation}
\label{2eq5} S_{q}=-\tau_{\rm eff}\int dtdrd^{p}x\sqrt{-{\rm
det}(g_{ab}+2\pi\alpha^{\prime}F_{ab})},
\end{equation}
where $\tau_{\rm eff}$ is given by
$$\tau_{\rm eff}=T_{q}L^{n}{\rm Vol}({S^{n}})e^{-\phi_{0}}.$$
However, as pointed out in~\cite{Hartnoll:2009ns}, incorporating a
non-trivial dilaton might be helpful to realistic model-building. So
here we will also discuss the situation with non-trivial dilaton,
whose effective action turns out to be
\begin{equation}
\label{2eq6} \tilde{S}_{q}=-\tilde{\tau}_{\rm eff}\int
dtdrd^{p}xe^{-\phi}\sqrt{-{\rm
det}(g_{ab}+2\pi\alpha^{\prime}F_{ab})},
\end{equation}
where
$$\tilde{\tau}_{\rm eff}=T_{q}L^{n}{\rm Vol}({S^{n}}).$$

In the following sections we will mainly focus on the details for
massless charge carriers, both with trivial and non-trivial dilaton.
Remarks on the corresponding properties for massive charge carriers
will be given in section 6. For simplicity, we will set $L=1$ in the
following and restore the factor of $L$ in the final results for the
physical quantities, such as the conductivity, by dimensional
analysis.

\section{Thermodynamics of massless charge carriers }
We discuss the thermodynamics of massless charge carriers at finite
charge density in this section. As claimed
in~\cite{Kobayashi:2006sb}, once a non-vanishing worldvolume gauge
field strength $F_{0r}$ is turned on, only black hole embeddings are
physically allowed. That is, the internal $S^{n}$ of the probe
D$q$-brane never collapse to zero volume and the D$q$-brane extends
to, and intersects the horizon. However, in~\cite{Nakamura:2006xk,
Nakamura:2007nx} it was argued that Minkowski embeddings were
physical and should be considered, due to the fact that the black
hole configurations cannot realize chemical potential below a
critical value. Later, this problem was resolved
in~\cite{Karch:2007br} by including Minkowski branes with a constant
$A_{0}$.

In this section we will divide both sides of~(\ref{2eq5})
and~(\ref{2eq6}) by the volume of $\mathbb{R}^{1,p}$, working with
action densities. We also turn on the time component of the
worldvolume gauge field $A_{0}$ and absorb the factor of
$2\pi\alpha^{\prime}$ into $A_{0}$ for simplicity. Then the DBI
action can be expressed as
\begin{equation}
S_{q}=-\tau_{\rm eff}\int drr^{p}\sqrt{r^{2z-2}-A^{\prime2}_{0}}.
\end{equation}
By defining the charge density
\begin{equation}
\rho=\frac{\delta\mathcal{L}}{\delta A^{\prime}_{0}}=\frac{\tau_{\rm
eff}r^{p}A^{\prime}_{0}}{\sqrt{r^{2z-2}-A^{\prime2}_{0}}},
\end{equation}
the solution for $A^{\prime}_{0}$ is given by
\begin{equation}
A^{\prime}_{0}=\frac{dr^{z-1}}{\sqrt{r^{2p}+d^{2}}},
\end{equation}
where $d=\rho/\tau_{\rm eff}$.

In the grand-canonical ensemble, the free energy density $\Omega$ is
given by minus the on-shell value of the action. After plugging in
the solution for $A^{\prime}_{0}$ we obtain
\begin{equation}
\Omega=\tau_{\rm
eff}\int^{\infty}_{r_{+}}\frac{r^{z-1+2p}}{\sqrt{r^{2p}+d^{2}}}dr.
\end{equation}
The divergence can be regulated by background subtraction or local
counterterms. The chemical potential $\mu$ in the grand-canonical
ensemble is determined by $A_{0}(\infty)$. Notice that
$A_{0}(r_{+})=0$, so the chemical potential is given by
\begin{equation}
\mu=\int^{\infty}_{r_{+}}A^{\prime}_{0}dr.
\end{equation}
It can be seen that our expressions above reduce to those derived
in~\cite{Karch:2009eb} when $z=1$.

It was observed in~\cite{Karch:2007br} that such integrals can be
worked out in terms of Beta functions and incomplete Beta functions,
whose definitions are given as follows
\begin{equation}
B(a,b)=\frac{\Gamma(a)\Gamma(b)}{\Gamma(a+b)}=\int^{1}_{0}dt(1-t)^{b-1}t^{a-1}
=\int^{\infty}_{0}du(1+u)^{-(a+b)}u^{a-1},
\end{equation}
\begin{equation}
B(x;a,b)=\int^{x}_{0}(1-t)^{b-1}t^{a-1}=\int^{x/(1-x)}_{0}du(1+u)^{-(a+b)}u^{a-1}.
\end{equation}
Therefore one can arrive at the analytic expressions for the free
energy and the chemical potential
\begin{equation}
\label{3eq8}
\mu=\mu_{0}-\frac{r^{z}_{+}}{z}{}_{2}F_{1}[\frac{z}{2p},\frac{1}{2};
1+\frac{z}{2p};-\frac{r^{2p}_{+}}{d^{2}}],
\end{equation}
\begin{equation}
\label{3eq9} \Omega=\Omega_{0}-\frac{\tau_{\rm
eff}r^{2p+z}_{+}}{(2p+z)d}{}_{2}F_{1}[1+\frac{z}{2p},
\frac{1}{2};2+\frac{z}{2p};-\frac{r^{2p}_{+}}{d^{2}}]+\Omega_{\rm
ct},
\end{equation}
where the following formulae have been used
\begin{equation}
B(x;a,b)=a^{-1}x^{a}{}_{2}F_{1}[a,1-b;a+1;x],
\end{equation}
\begin{equation}
{}_{2}F_{1}[a,b;c;x]=(1-x)^{-a}{}_{2}F_{1}[a,c-b;c;\frac{x}{x-1}].
\end{equation}
The parameters $\mu_{0}$ and $\Omega_{0}$ in~(\ref{3eq8})
and~(\ref{3eq9}) are zero-temperature values, given by
\begin{equation}
\mu_{0}=d^{z/p}\alpha(p),~~~\alpha(p)=\frac{1}{2z}
B(1+\frac{z}{2p},\frac{1}{2}-\frac{z}{2p}),
\end{equation}
\begin{equation}
\Omega_{0}=-\frac{z\tau_{\rm
eff}}{(z+p)\alpha(p)^{p/z}}\mu_{0}^{1+p/z}.
\end{equation}
One can check that these results are in agreement with those
obtained in~\cite{Karch:2008fa} when $z=1$.

Next we calculate the thermodynamic quantities in the
grand-canonical ensemble. It should be pointed out that the term
$\Omega_{\rm ct}$ in~(\ref{3eq9}) stands for the contribution from
the counterterms in the spirit of holographic renormalization. As
was observed in~\cite{Karch:2008fa}, $\Omega_{\rm ct}$ is
independent of the density. Since we do not have a well-established
holographic renormalization scheme for probe D-branes in Lifshitz
backgrounds, we assume that here the contribution from the
counterterms is still independent of the density. We shall focus on
the density-dependent part of the free energy
$\Delta\Omega\equiv\Omega-\Omega_{\rm ct}$. Notice that at low
temperature, both~(\ref{3eq8}) and~(\ref{3eq9}) can be expanded as
series in $T/\mu_{0}\ll1$. The charge density is
\begin{equation}
\rho=-\frac{\partial\Delta\Omega}{\partial\mu}=\tau_{\rm eff}d,
\end{equation}
which provides a consistency check. One then computes the entropy
density $s(\mu, T)$ in the grand-canonical ensemble
\begin{equation}
\label{3eq15} s(\mu, T)=-\big(\frac{\partial\Delta\Omega}{\partial
T}\big)_{\mu} =\frac{4\pi\rho}{\lambda}+\frac{\tau_{\rm
eff}}{2zd}(\frac{4\pi z}{\lambda})^{1+2p/z}T^{2p/z},
\end{equation}
where the first term gives the entropy density at zero temperature.
Finally, the specific heat at low temperature is given by
\begin{equation}
\label{3eq16} c_{V}=T\big(\frac{\partial s}{\partial T}\big)_{\rho}
=\frac{\tau_{\rm eff}p}{z^{2}d}(\frac{4\pi
z}{\lambda})^{1+2p/z}T^{2p/z}.
\end{equation}
One can find agreement with the results obtained
in~\cite{Karch:2008fa} once again in the $z=1$ case.

One may compare the specific heat $c_{V}$ in~(\ref{3eq16}) to that
of a gas of free bosons or fermions. As is well known, the
low-temperature specific heat for a gas of free bosons is
proportional to $T^{p}$, while the low-temperature specific heat for
a gas fermions is proportional to $T$ for any $p$. Therefore our
result indicates that when $z=2$, the behavior of the specific heat
looks like a gas of free bosons. On the other hand, it resembles a
gas of fermions when $z=2p$. Finally, it may be regarded as a new
type of quantum liquid in other cases.

Another interesting quantity is the speed of sound at low
temperature. In this limit the pressure is given by
\begin{equation}
P=-\Omega_{0}=\frac{z\tau_{\rm
eff}}{(z+p)\alpha(p)^{p/z}}\mu_{0}^{1+p/z}.
\end{equation}
The energy density is
\begin{equation}
\epsilon=\Omega_{0}+\mu_{0}\rho=\frac{p\tau_{\rm
eff}}{(z+p)\alpha(p)^{p/z}}\mu_{0}^{1+p/z}.
\end{equation}
One can easily find that $\epsilon=(p/z)P$. Thus the speed of sound
is determined by
\begin{equation}
c^{2}_{s}=\frac{dP}{d\epsilon}=\frac{z}{p}.
\end{equation}
Note that an upper bound on the speed of sound was proposed
in~\cite{Hohler:2009tv, Cherman:2009tw}, which gives
$c^{2}_{s}\leq1/3$ in five-dimensional bulk spacetime. Then such a
bound might be violated in five dimensions, i.e. $p=3$, when $z>1$.
However, it is believed that such a bound always holds at high
temperature. Therefore the violation of this bound with $z>1$ may
not be seen as a contradiction.

When a non-trivial dilaton field, which behaves as $e^{-\phi}\propto
r^{\kappa}$, is incorporated, the charge density, chemical potential
and on-shell action turn out to be
\begin{equation}
\tilde{\rho}=\frac{\delta\mathcal{L}}{\delta
A^{\prime}_{0}}\sim\frac{\tilde{\tau}_{\rm
eff}r^{p+\kappa}A^{\prime}_{0}}{\sqrt{r^{2z-2}-A^{\prime2}_{0}}},
\end{equation}
\begin{equation}
\tilde{\mu}=\int^{\infty}_{r_{+}}A^{\prime}_{0}dr\sim\int^{\infty}_{r_{+}}
\frac{\tilde{d}r^{z-1}}{\sqrt{r^{2(p+\kappa )}+\tilde{d}^{2}}}dr,
\end{equation}
\begin{equation}
\tilde{\Omega}\sim\tilde{\tau}_{\rm
eff}\int^{\infty}_{r_{+}}\frac{r^{z-1+2(p+\kappa)}}{\sqrt{r^{2(p+\kappa)}
+\tilde{d}^{2}}}dr,
\end{equation}
where $\tilde{\tau}_{\rm eff}=T_{q}L^{n}{\rm Vol}({S^{n}})$ and
$\tilde{d}=\tilde{\rho}/\tilde{\tau}_{\rm eff}$. It can be seen that
all the thermodynamic quantities in the presence of a non-trivial
dilaton can be obtained by simply replacing $p$ by $p+\kappa$ in the
expressions for those with a constant dilaton.

In particular, the specific heat becomes
\begin{equation}
\tilde{c}_{V}\sim T^{2(p+\kappa)/z}.
\end{equation}
Therefore the specific heat scales as a gas of free bosons when
$z=2(p+\kappa)/p$ and as a gas of fermions when $z=2(p+\kappa)$. It
may be regarded as a new type of quantum liquid for other cases. The
speed of sound is given by
\begin{equation}
\tilde{c}^{2}_{s}=\frac{z}{p+\kappa}.
\end{equation}
Thus the upper bound proposed in~\cite{Hohler:2009tv,
Cherman:2009tw} can be saturated by setting $z=(p+\kappa)/3$ in five
dimensions.
\section{Conductivity and resistivity}
We calculate the DC conductivity, DC Hall conductivity and
resistivity in this section, following the method proposed
in~\cite{Karch:2007pd} and~\cite{O'Bannon:2007in}. We find that in
the large charge density limit, the DC conductivity exhibits certain
universal behavior, independent of the spacetime dimension. On the
other hand, the DC Hall conductivity always takes the same form as
the four-dimensional counterpart. Finally, we obtain the resistivity
in different limits and figure out the requirements of linear
dependence on the temperature $T$.
\subsection{DC conductivity}
Generally speaking, there are two different ways for calculating DC
conductivity in flavor brane systems. One is the traditional way,
that is, to obtain the conductivity via Kubo formula. The other
elegant way, which was proposed in~\cite{Karch:2007pd}, is more
efficient for flavor brane systems. The main strategy is to turn on
an electric field $E\equiv F_{tx}$ on the D-brane probe and compute
the corresponding current $J^{x}$ in the boundary theory. Then the
conductivity can be read off from Ohm's law $J^{x}=\sigma E$.

We will calculate the DC conductivity by applying the second method
in this section and present discussions on the first method in the
next section. To make comparison with~\cite{Hartnoll:2009ns}, we
take the following coordinate transformation before performing the
calculations
\begin{equation}
v=\frac{1}{r},~~~v_{+}=\frac{1}{r_{+}}.
\end{equation}
Then the metric of the $(p+2)$-dimensional Lifshitz black hole
becomes(setting $L=1$)
\begin{equation}
ds^{2}=-\frac{f(v)}{v^{2z}}dt^{2}+\frac{dv^{2}}{v^{2}f(v)}+\frac{1}{v^{2}}d\vec{x}^{2}_{p}
\end{equation}
and the temperature is given by
\begin{equation}
T=\frac{\lambda}{4\pi zv^{z}_{+}}.
\end{equation}

After taking the ansatz for the worldvolume gauge field
\begin{equation}
A=A_{0}(v)dt+(-Et+h(v))dx,
\end{equation}
the effective action~(\ref{2eq5}) becomes
\begin{equation}
\label{4eq5} S_{q}=-\tau_{\rm eff}\int
dtdvd^{p}xg_{xx}^{\frac{p-1}{2}}\sqrt{-g_{tt}g_{vv}g_{xx}-(
2\pi\alpha^{\prime})^2(g_{vv}E^{2}+g_{xx}A^{\prime2}_{0}+g_{tt}h^{\prime2})}.
\end{equation}
Note that the action contains only $A^{\prime}_{0}$ and
$h^{\prime}(v)$, which leads to two conserved quantities
\begin{equation}
\label{4eq6}
C=-\frac{g^{\frac{p+1}{2}}_{xx}A^{\prime}_{0}}{\sqrt{-g_{tt}g_{vv}g_{xx}-(
2\pi\alpha^{\prime})^2(g_{vv}E^{2}+g_{xx}A^{\prime2}_{0}+g_{tt}h^{\prime2})}},
\end{equation}
\begin{equation}
\label{4eq7}
H=-\frac{g^{\frac{p-1}{2}}_{xx}g_{tt}h^{\prime}}{\sqrt{-g_{tt}g_{vv}g_{xx}-(
2\pi\alpha^{\prime})^2(g_{vv}E^{2}+g_{xx}A^{\prime2}_{0}+g_{tt}h^{\prime2})}}.
\end{equation}
It can be observed that
\begin{equation}
\label{4eq8} g_{tt}h^{\prime}C=g_{xx}A^{\prime}_{0}H.
\end{equation}

By combining~(\ref{4eq6}),~(\ref{4eq7}) and~(\ref{4eq8}), we can
arrive at the following expression for $A^{\prime}_{0}$
\begin{equation}
\label{4eq9}
A^{\prime2}_{0}=-\frac{C^{2}g_{tt}g_{vv}}{g_{xx}}\frac{g_{tt}g_{xx}
+(2\pi\alpha^{\prime2}E^{2})}{g_{tt}g_{xx}^{p}+(2\pi\alpha^{\prime})^{2}(C^{2}g_{tt}+H^{2}g_{xx})},
\end{equation}
which results in the asymptotic behavior for $A_{0}$ near the
boundary $v~\rightarrow~0$
\begin{equation}
A_{0}=\mu-\frac{C}{z-p}\frac{1}{v^{z-p}}+\cdots,~~~{\rm for}~z\neq
p,
\end{equation}
or
\begin{equation}
A_{0}=\mu+C\log\frac{v}{\Lambda}+\cdots,~~~{\rm for}~z=p,
\end{equation}
where $\mu$ denotes the chemical potential and $\Lambda$ stands for
the UV cutoff. In the meantime, the asymptotic behavior of $h(v)$ is
given by
\begin{equation}
h(v)=h_{0}+\frac{H}{(z+p-2)}v^{z+p-2}+\cdots,
\end{equation}
where we can set $h_{0}=0$.

After substituting~(\ref{4eq8}) and~(\ref{4eq9}) back
into~(\ref{4eq5}), the effective action becomes
\begin{equation}
\label{4eq13} S_{q}=-\tau_{\rm eff}\int
dtdvd^{p}xg_{xx}^{\frac{2p-1}{2}}\sqrt{-g_{tt}g_{vv}}\sqrt{\frac{g_{tt}g_{xx}
+(2\pi\alpha^{\prime})^{2}E^{2}}{g_{tt}g_{xx}^{p}+(2\pi\alpha^{\prime})^{2}(C^{2}g_{tt}+
H^{2}g_{xx})}}.
\end{equation}
As pointed out in~\cite{Karch:2007pd}, both the numerator and the
denominator of the term inside the square root of~(\ref{4eq13})
change sign between the horizon $v=v_{+}$ and the boundary $v=0$.
Therefore in order to ensure the reality of the on-shell action, the
sign change must appear at the same radial position
$0<v_{\ast}<v_{+}$,
\begin{equation}
\label{4eq14} g_{tt}g_{xx}
+(2\pi\alpha^{\prime})^{2}E^{2}\big|_{v=v_{\ast}}=0,
\end{equation}
\begin{equation}
\label{4eq15}
g_{tt}g_{xx}^{p}+(2\pi\alpha^{\prime})^{2}(C^{2}g_{tt}+
H^{2}g_{xx})\big|_{v=v_{\ast}}=0.
\end{equation}

From~(\ref{4eq14}) we can obtain
\begin{equation}
\label{4eq16}
f(v_{\ast})=(2\pi\alpha^{\prime})^{2}E^{2}v^{2z+2}_{\ast}.
\end{equation}
Finally, by substituting~(\ref{4eq16}) into~(\ref{4eq15}) and making
the following identifications for the currents
\begin{equation}
J^{t}=(2\pi\alpha^{\prime})^{2}\tau_{\rm
eff}C,~~~J^{x}=(2\pi\alpha^{\prime})^{2}\tau_{\rm eff}H,
\end{equation}
we can read off the conductivity
\begin{equation}
\label{4eq18} \sigma=\sqrt{(2\pi\alpha^{\prime})^{4}\tau^{2}_{\rm
eff}(\frac{L}{v_{\ast}})^{2p-4}+(\frac{2\pi\alpha^{\prime}}{L^{2}})^{2}(J^{t})^{2}v^{4}_{\ast}}
\end{equation}
from Ohm's law, where we have restored the factor of $L$ by
dimensional analysis. Notice that the right hand side
of~(\ref{4eq18}) is still written as the mean square root of two
terms, which is similar to the result for general D$p$/D$q$ systems
in~\cite{Karch:2007pd}. The first term may be interpreted as
contribution from thermally produced pairs of charge carriers, which
is expected to be Boltzmann suppressed when the mass of the charge
carriers becomes large. The second term is independent of the
spacetime dimension. Furthermore, (\ref{4eq18}) reduces to the
four-dimensional result obtained in~\cite{Hartnoll:2009ns} when
$p=2$.

When a non-trivial dilaton is introduced, the effective action
becomes
\begin{equation}
\tilde{S}_{q}=-\tilde{\tau}_{\rm eff}\int
dtdvd^{p}xe^{-\phi}g_{xx}^{\frac{p-1}{2}}\sqrt{-g_{tt}g_{vv}g_{xx}-(
2\pi\alpha^{\prime})^2(g_{vv}E^{2}+g_{xx}A^{\prime2}_{0}+g_{tt}h^{\prime2})}.
\end{equation}
The corresponding two conserved quantities are given by
\begin{equation}
\tilde{C}=-\frac{e^{-\phi}g^{\frac{p+1}{2}}_{xx}A^{\prime}_{0}}{\sqrt{-g_{tt}g_{vv}g_{xx}-(
2\pi\alpha^{\prime})^2(g_{vv}E^{2}+g_{xx}A^{\prime2}_{0}+g_{tt}h^{\prime2})}},
\end{equation}
\begin{equation}
\tilde{H}=-\frac{e^{-\phi}g^{\frac{p-1}{2}}_{xx}g_{tt}h^{\prime}}{\sqrt{-g_{tt}g_{vv}g_{xx}-(
2\pi\alpha^{\prime})^2(g_{vv}E^{2}+g_{xx}A^{\prime2}_{0}+g_{tt}h^{\prime2})}}.
\end{equation}
Now the on-shell action turns out to be
\begin{equation}
\tilde{S}_{q}=-\tilde{\tau}_{\rm eff}\int
dtdvd^{p}xe^{-2\phi}g_{xx}^{\frac{2p-1}{2}}\sqrt{-g_{tt}g_{vv}}\sqrt{\frac{g_{tt}g_{xx}
+(2\pi\alpha^{\prime})^{2}E^{2}}{e^{-2\phi}g_{tt}g_{xx}^{p}+(2\pi\alpha^{\prime})^{2}(C^{2}g_{tt}+
H^{2}g_{xx})}}.
\end{equation}
Finally we can obtain the DC conductivity in a similar way
\begin{equation}
\label{4eq23}
\tilde{\sigma}=\sqrt{(2\pi\alpha^{\prime})^{4}e^{-2\phi_{\ast}}\tilde{\tau}^{2}_{\rm
eff}(\frac{L}{v_{\ast}})^{2p-4}+(\frac{2\pi\alpha^{\prime}}{L^{2}})^{2}(J^{t})^{2}v^{4}_{\ast}},
\end{equation}
which reduces to~(\ref{4eq18}) in the limit of $\phi=\phi_{0}$.
\subsection{DC Hall conductivity}
One can also calculate the DC Hall conductivity by applying the
techniques of~\cite{Karch:2007pd}. For general D$p$/D$q$ systems
such analysis was carried out in~\cite{O'Bannon:2007in} and the Hall
conductivity in four-dimensional Lifshitz background was obtained
in~\cite{Hartnoll:2009ns}. Here we extend the analysis to general
$p+2$-dimensional spacetime and we will see that the Hall
conductivity exhibits universal behavior.

Once we introduce a constant magnetic field on the worldvolume of
the probe D-brane, the ansatz for the $U(1)$ gauge field becomes
\begin{equation}
\label{4eq24}
A_{0}=A_{0}(v),~~~A_{x}(v,t)=-Et+f_{x}(v),~~~A_{y}(v,x)=Bx+f_{y}(v).
\end{equation}
Then the effective action turns out to be
\begin{eqnarray}
S_{q}&=&-\tau_{\rm eff}\int dtdvd^{p}x\sqrt{-{\rm
det}(g_{ab}+2\pi\alpha^{\prime}F_{ab})}\nonumber\\
&\equiv&-\tau_{\rm eff}\int
dtdvd^{p}xg_{xx}^{\frac{p-2}{2}}\sqrt{GF},
\end{eqnarray}
where
\begin{eqnarray}
\label{4eq26}
GF&=&-g_{tt}g_{vv}g_{xx}^{2}-(2\pi\alpha^{\prime})^{2}g_{xx}^{2}A_{0}^{\prime2}
-(2\pi\alpha^{\prime})^{2}g_{vv}g_{xx}E^{2}\nonumber\\
&
&-(2\pi\alpha^{\prime})^{2}g_{tt}g_{vv}B^{2}-(2\pi\alpha^{\prime})^{2}g_{tt}g_{xx}
(f_{x}^{\prime2}+f_{y}^{\prime2})\nonumber\\
&
&-(2\pi\alpha^{\prime})^{4}A_{0}^{\prime2}B^{2}-(2\pi\alpha^{\prime})^{4}f_{y}^{\prime2}E^{2}
+2(2\pi\alpha^{\prime})^{4}A_{0}^{\prime}f_{y}^{\prime}EB.
\end{eqnarray}
Since the action depends on the derivatives $A_{0}^{\prime},
f_{x}^{\prime}$ and $f_{y}^{\prime}$, we can obtain the following
conserved quantities
\begin{equation}
[-g_{xx}^{2}A_{0}^{\prime}-(2\pi\alpha^{\prime})^{2}B^{2}A_{0}^{\prime}
+(2\pi\alpha^{\prime})^{2}EBf_{y}^{\prime}](\sqrt{GF})^{-1}g_{xx}^{\frac{p-2}{2}}=C,
\end{equation}
\begin{equation}
-g_{tt}g_{xx}f_{x}^{\prime}(\sqrt{GF})^{-1}g_{xx}^{\frac{p-2}{2}}=H,
\end{equation}
\begin{equation}
[-g_{tt}g_{xx}f_{y}^{\prime}-(2\pi\alpha^{\prime})^{2}E^{2}f_{y}^{\prime}
+(2\pi\alpha^{\prime})^{2}EBA_{0}^{\prime}](\sqrt{GF})^{-1}g_{xx}^{\frac{p-2}{2}}=M.
\end{equation}
After solving for $A_{0}^{\prime}, f_{x}^{\prime}$ and
$f_{y}^{\prime}$ from the above equations and plugging the solutions
back into the effective action, we obtain
\begin{equation}
S_{q}=\tau_{\rm eff}\int
dvg_{xx}^{p-1}\sqrt{-g_{tt}g_{vv}}\frac{\xi}{\sqrt{\xi\eta-a^{2}}},
\end{equation}
where
\begin{equation}
\label{4eq31}
\xi=-[(2\pi\alpha^{\prime})^{2}E^{2}g_{xx}+(2\pi\alpha^{\prime})
^{2}B^{2}g_{tt}+g_{tt}g_{xx}^{2}],
\end{equation}
\begin{equation}
\label{4eq32}
\eta=-g_{tt}g_{xx}^{p}-(2\pi\alpha^{\prime})^{2}[g_{tt}C^{2}+g_{xx}(H^{2}+M^{2})],
\end{equation}
\begin{equation}
\label{4eq33} a=(2\pi\alpha^{\prime})^{2}(MEg_{xx}-BCg_{tt}).
\end{equation}

As pointed out in~\cite{O'Bannon:2007in}, to ensure the reality of
the effective action, $\xi, \eta$ and $a$ must share the same zero
$v_{\ast}$. From~(\ref{4eq31}) we obtain
\begin{equation}
\label{4eq34}
f(v_{\ast})=\frac{(2\pi\alpha^{\prime})^{2}E^{2}v_{\ast}^{2z+2}}{1+
(2\pi\alpha^{\prime})^{2}B^{2}v_{\ast}^{4}}.
\end{equation}
Then we can substitute~(\ref{4eq34}) into~(\ref{4eq32})
and~(\ref{4eq33}) to solve for $H$ and $M$, and make the following
identifications for the currents
\begin{equation}
J^{t}=(2\pi\alpha^{\prime})^{2}\tau_{\rm
eff}C,~~~J^{x}=(2\pi\alpha^{\prime})^{2}\tau_{\rm
eff}H,~~~J^{y}=(2\pi\alpha^{\prime})^{2}\tau_{\rm eff}M.
\end{equation}
Finally, the conductivity tensor $J^{i}=\sigma^{ij}E_{j}$ are given
by
\begin{equation}
\sigma^{xx}=\frac{1}{1+(\frac{2\pi\alpha^{\prime}}{L^{2}})^{2}B^{2}v^{4}_{\ast}}
\sqrt{(2\pi\alpha^{\prime})^{4}\tau_{\rm
eff}^{2}(\frac{L}{v_{\ast}})^{2p-4}(1+(\frac{2\pi\alpha^{\prime}}{L^{2}})^{2}
B^{2}v_{\ast}^{4})+(\frac{2\pi\alpha^{\prime}}{L^{2}})^{2}(J^{t})^{2}v_{\ast}^{4}},
\end{equation}
\begin{equation}
\sigma^{xy}=\frac{BJ^{t}v_{\ast}^{4}}{(\frac{L^{2}}{2\pi\alpha^{\prime}})^{2}+B^{2}v_{\ast}^{4}}.
\end{equation}
Here are several remarks on the conductivity tensor
\begin{itemize}
\item $\sigma^{xx}$ reduces to the expression obtained in previous
section when $B=0$.
\item $\sigma^{xy}$ does not depend on the dimension of spacetime,
which is the same as that for general D$p$/D$q$ systems analyzed
in~\cite{O'Bannon:2007in}.
\item One interesting quantity for strange metals is the ratio
$\sigma^{xx}/\sigma^{xy}$. It has been known that the ratio scales
as $\sigma^{xx}/\sigma^{xy}\sim T^{2}$ for strange metals, while
$\sigma^{xx}/\sigma^{xy}\sim1/\sigma^{xx}$ in Drude theory. It can
be easily seen that
\begin{equation}
\frac{\sigma^{xx}}{\sigma^{xy}}=(\frac{L^{2}}{2\pi\alpha^{\prime}})^{2}
\frac{1}{J^{t}Bv_{\ast}^{4}}\sqrt{(2\pi\alpha^{\prime})^{4}\tau_{\rm
eff}^{2}(\frac{L}{v_{\ast}})^{2p-4}(1+(\frac{2\pi\alpha^{\prime}}{L^{2}})^{2}
B^{2}v_{\ast}^{4})+(\frac{2\pi\alpha^{\prime}}{L^{2}})^{2}(J^{t})^{2}v_{\ast}^{4}}.
\end{equation}
When the second term in the square root dominates and $B$ is small,
which is relevant to experimental limit, the ratio becomes
$\sigma^{xx}/\sigma^{xy}\sim v_{\ast}^{-2}\sim1/\sigma^{xx}$. Thus
the result obtained from the probe calculation mimics the Drude
theory in arbitrary dimensions.
\end{itemize}

When a non-trivial dilaton is introduced, the effective action
becomes
\begin{equation}
\tilde{S}_{q}=-\tilde{\tau}_{\rm eff}\int
dtdvd^{p}xe^{-\phi}\sqrt{-{\rm
det}(g_{ab}+2\pi\alpha^{\prime}F_{ab})}.
\end{equation}
The ansatz for the worldvolume gauge fields is still given
by~(\ref{4eq24}). Therefore the effective action turns out to be
\begin{equation}
\tilde{S}_{q}=-\tilde{\tau}_{\rm eff}\int
dtdvd^{p}xe^{-\phi}g_{xx}^{\frac{p-2}{2}}\sqrt{GF},
\end{equation}
where $GF$ is still written in the form of~(\ref{4eq26}). The
corresponding conserved quantities are given by
\begin{equation}
[-g_{xx}^{2}A_{0}^{\prime}-(2\pi\alpha^{\prime})^{2}B^{2}A_{0}^{\prime}
+(2\pi\alpha^{\prime})^{2}EBf_{y}^{\prime}](\sqrt{GF})^{-1}e^{-\phi
}g_{xx}^{\frac{p-2}{2}}=\tilde{C},
\end{equation}
\begin{equation}
-g_{tt}g_{xx}f_{x}^{\prime}(\sqrt{GF})^{-1}e^{-\phi}g_{xx}^{\frac{p-2}{2}}=\tilde{H},
\end{equation}
\begin{equation}
[-g_{tt}g_{xx}f_{y}^{\prime}-(2\pi\alpha^{\prime})^{2}E^{2}f_{y}^{\prime}
+(2\pi\alpha^{\prime})^{2}EBA_{0}^{\prime}](\sqrt{GF})^{-1}
e^{-\phi}g_{xx}^{\frac{p-2}{2}}=\tilde{M}.
\end{equation}
The on-shell effective action can be written in the following form
\begin{equation}
\tilde{S}_{q}=\tilde{\tau}_{\rm eff}\int
dve^{-2\phi}g_{xx}^{p-1}\sqrt{-g_{tt}g_{vv}}\frac{\xi}{\sqrt{\xi\tilde{\eta}-a^{2}}},
\end{equation}
where $\xi$ and $a$ are still given by~(\ref{4eq31})
and~(\ref{4eq33}) with $C, H$ and $M$ replaced by $\tilde{C},
\tilde{H}$ and $\tilde{M}$, while $\tilde{\eta}$ is slightly
different from $\eta$,
\begin{equation}
\tilde{\eta}=-g_{tt}g_{xx}^{p}e^{-2\phi}
-(2\pi\alpha^{\prime})^{2}[g_{tt}\tilde{C}+g_{xx}(\tilde{H}+\tilde{M})].
\end{equation}
Finally the conductivity tensor is given by
\begin{equation}
\tilde{\sigma}^{xx}=\frac{1}{1+(\frac{2\pi\alpha^{\prime}}{L^{2}})^{2}B^{2}v^{4}_{\ast}}
\sqrt{(2\pi\alpha^{\prime})^{4}e^{-2\phi_{\ast}}\tilde{\tau}_{\rm
eff}^{2}(\frac{L}{v_{\ast}})^{2p-4}(1+(\frac{2\pi\alpha^{\prime}}{L^{2}})^{2}
B^{2}v_{\ast}^{4})+(\frac{2\pi\alpha^{\prime}}{L^{2}})^{2}(J^{t})^{2}v_{\ast}^{4}},
\end{equation}
\begin{equation}
\sigma^{xy}=\frac{BJ^{t}v_{\ast}^{4}}{(\frac{L^{2}}{2\pi\alpha^{\prime}})^{2}+B^{2}v_{\ast}^{4}}.
\end{equation}
It can be easily seen that $\tilde{\sigma}^{xx}$ reduces to
$\sigma^{xx}$ when $\phi=\phi_{0}$ and $\sigma^{xy}$ remains
invariant.
\subsection{Resistivity in different limits}
Once we have obtained the conductivity, we may investigate the
behavior of the resistivity to see if it exhibits strange metallic
behavior, that is, the resistivity has a linear dependence on the
temperature $T$. Since the conductivity contains two terms inside
the square root, it is more transparent to study the resistivity in
two different limits.

Recall that the conductivity is given by
$$\sigma=\sqrt{(2\pi\alpha^{\prime})^{4}\tau^{2}_{\rm
eff}(\frac{L}{v_{\ast}})^{2p-4}+(\frac{2\pi\alpha^{\prime}}{L^{2}})^{2}(J^{t})^{2}v^{4}_{\ast}}.
$$
One limit which is experimentally interesting is large $J^{t}$, then
the first term is subdominant and the conductivity turns out to be
\begin{equation}
\sigma=\frac{2\pi\alpha^{\prime}}{L^{2}}J^{t}v^{2}_{\ast}.
\end{equation}
Then by combining the fact that $T\propto1/v^{z}_{+}$ we can obtain
\begin{equation}
\label{4eq49} \rho=\frac{1}{\sigma}\sim\frac{T^{2/z}}{J^{t}}.
\end{equation}
It seems that the resistivity exhibit a universal behavior in the
limit of large charge density, that is, it is proportional to
$T^{2/z}$ and is independent of the spacetime dimension. Therefore
the conductivity is of the strange metal type in all dimensions for
$z=2$.

The other opposite limit is the zero density limit, which means that
the first term in the square root dominates. Then the conductivity
is given by
\begin{equation}
\sigma=(2\pi\alpha^{\prime})\tau_{\rm
eff}(\frac{L}{v_{\ast}})^{p-2}.
\end{equation}
When $p=2$, the conductivity becomes constant, which agrees with the
conclusion in~\cite{Herzog:2007ij}. For the $p>2$ cases, the
resistivity becomes
\begin{equation}
\rho\sim T^{-(p-2)/z},
\end{equation}
where we have used the fact that $T\sim1/v^{z}_{+}$. One can see
that for $p>2$, the linear dependence on the temperature cannot be
realized for $z>0$.

When a non-trivial dilaton is incorporated, the conductivity reads
$$\tilde{\sigma}=\sqrt{(2\pi\alpha^{\prime})^{4}e^{-2\phi_{\ast}}\tilde{\tau}^{2}_{\rm
eff}(\frac{L}{v_{\ast}})^{2p-4}+(\frac{2\pi\alpha^{\prime}}{L^{2}})^{2}(J^{t})^{2}v^{4}_{\ast}}.$$
One can easily observe that the resistivity is still given
by~(\ref{4eq49}) in the large charge density limit, that is, the
resistivity has a linear dependence on $T$ with $z=2$ in all
dimensions. When the first term dominates, we can assume that
$e^{-\phi}\propto1/v^{\kappa}$, then
\begin{equation}
\rho\sim T^{-\frac{\kappa+p-2}{z}}.
\end{equation}
Therefore the linear dependence on the temperature can be realized
by requiring $z=-(\kappa+p-2)$. In particular, in four-dimensional
spacetime with $p=2$, we can arrive at the linear dependence as long
as $z=-\kappa$.
\section{DC conductivity revisited and charge diffusion constant}
We obtained the DC conductivity in the previous section by
introducing an electric field $E$ on the worldvolume of the probe
D-branes and making use of Ohm's law $J=\sigma E$. In some sense
this approach can be seen as ``macroscopic'', while the
corresponding ``microscopic'' approach is to calculate the
conductivity via Kubo formula. A systematic analysis for calculating
the holographic spectral functions for D$p$/D$q$ systems at finite
chemical potential and spatial momentum was carried out
in~\cite{Mas:2008jz}, where it was found that the conductivity
obtained via Kubo formula was in agreement with that obtained in the
``macroscopic'' approach in the vanishing electric field limit. This
can be seen as a non-trivial check of consistency. Holographic
spectral functions with non-vanishing electric field strength was
investigated in~\cite{Mas:2009wf}, where precise agreement was found
once again. Furthermore, conductivity and diffusion constant for
D$p$/D$q$ systems were studied in~\cite{Mas:2008qs}, generalizing
the formulae derived via the membrane paradigm~\cite{Kovtun:2003wp}.
In this section we will revisit the conductivity
following~\cite{Mas:2008jz} and we will find that the result agrees
with that obtained in previous section. The charge diffusion
constant in such asymptotically Lifshitz backgrounds will also be
discussed.

\subsection{Calculating DC conductivity via Kubo formula}
Consider the following fluctuations of the worldvolume gauge field
\begin{equation}
A_{a}~\rightarrow~\delta_{a0}A_{0}(r)+\epsilon e^{-i(\omega
x^{0}-qx^{1})}\mathcal{A}_{a}(r),
\end{equation}
the DBI Lagrangian can be expanded in powers of $\epsilon$,
\begin{equation}
\mathcal{L}_{\rm
DBI}=\mathcal{L}_{0}+\epsilon\mathcal{L}_{1}+\epsilon^{2}\mathcal{L}_{2}+\cdots.
\end{equation}
Notice that $\mathcal{L}_{1}$ vanishes identically upon imposing the
equations of motion for the background fields. We shall not present
the detailed analysis for the fluctuations but just exhibit the
equations of motion,
\begin{equation}
\label{5eq3}
\mathcal{A}_{\perp}^{\prime\prime}+\partial_{r}\log[\sqrt{-\gamma}
\gamma^{rr}\gamma^{ii}]\mathcal{A}_{\perp}^{\prime}-\frac{\omega^{2}\gamma^{00}
+q^{2}\gamma^{ii}}{\gamma^{rr}}\mathcal{A}_{\perp}=0,
\end{equation}
\begin{equation}
\label{5eq4}
\mathcal{A}_{0}^{\prime\prime}+\partial_{r}\log[\sqrt{-\gamma}
\gamma^{rr}\gamma^{00}]\mathcal{A}_{0}^{\prime}-
q\frac{\gamma^{ii}}{\gamma^{rr}}(q\mathcal{A}_{0}+\omega\mathcal{A}_{1})=0,
\end{equation}
\begin{equation}
\label{5eq5}
\mathcal{A}_{1}^{\prime\prime}+\partial_{r}\log[\sqrt{-\gamma}
\gamma^{rr}\gamma^{ii}]\mathcal{A}_{1}^{\prime}-\omega\frac{\gamma^{00}}{\gamma^{rr}}
(q\mathcal{A}_{0}+\omega\mathcal{A}_{1})=0,
\end{equation}
as well as the constraint from the gauge choice $\mathcal{A}_{r}=0$
\begin{equation}
\label{5eq6}
-\omega\gamma^{00}\mathcal{A}^{\prime}_{0}+q\gamma^{ii}\mathcal{A}_{1}^{\prime}=0,
\end{equation}
for details see Appendix A of~\cite{Mas:2008jz}. Here we adopt the
notations of~\cite{Mas:2008jz},
\begin{equation}
F_{ab}=\partial_{a}A_{b}-\partial_{b}A_{a}\equiv
F^{(0)}_{ab}+\epsilon F^{(1)}_{ab},
\end{equation}
\begin{equation}
\gamma_{ab}=g_{ab}+F^{(0)}_{ab},~~~\gamma^{ab}=(\gamma_{ab})^{-1}.
\end{equation}
Notice that the equation of the transverse fluctuations
$\mathcal{A}_{\perp}$ is decoupled from others. Another point is
that~(\ref{5eq4}),~(\ref{5eq5}) and~(\ref{5eq6}) are not linearly
independent, as one can derive~(\ref{5eq5}) by
combining~(\ref{5eq4}) and~(\ref{5eq6}).

One may expect that we can perform the hydrodynamic expansions for
$\mathcal{A}_{a}$ and try to solve the corresponding equations, then
the retarded correlation functions can be obtained following the
standard procedure~\cite{Son:2002sd}. But it is quite difficult to
solve the relevant equations for the longitudinal fluctuations
indeed. However, the DC conductivity can be determined from the
retarded correlation function of the transverse modes via Kubo
formula
\begin{equation}
\sigma=\lim_{\omega\rightarrow0}{\rm
Im}\frac{1}{\omega}G^{R}_{\perp}(\omega, q=0).
\end{equation}
Then it is sufficient to work with $q=0$ to find the conductivity,
which was done for AdS case in~\cite{Kovtun:2008kx}. Under this
condition the equation of motion for $\mathcal{A}_{\perp}$ can be
simplified as
\begin{equation}
\mathcal{A}_{\perp}^{\prime\prime}+\partial_{r}\log[\sqrt{-\gamma}
\gamma^{rr}\gamma^{ii}]\mathcal{A}_{\perp}^{\prime}-\frac{\omega^{2}\gamma^{00}
}{\gamma^{rr}}\mathcal{A}_{\perp}=0,
\end{equation}
which allows us to find analytic solutions in the hydrodynamic
expansions. For general D$p$/D$q$ systems the final result was given
in~\cite{Mas:2008jz}
\begin{equation}
\label{5eq11}
\sigma=\mathcal{N}\sqrt{-\gamma\gamma^{00}\gamma^{rr}}\gamma^{ii}\big|_{r\rightarrow
r_{H}},
\end{equation}
where $\mathcal{N}=(2\pi\alpha^{\prime})^{2}\tau_{\rm eff}$ and
$r_{H}$ denotes the horizon. Notice that we have absorbed the
constant dilaton into $\mathcal{N}$.

For our specific model, the components of $\gamma_{ab}$ are given by
\begin{eqnarray}
\label{5eq12} &
&\gamma_{00}=g_{00}=-\frac{f(v)}{v^{2z}},~~~\gamma_{vv}=\frac{1}{v^{2}f(v)},
~~~\gamma_{ii}=g_{ii}=\frac{1}{v^{2}},\nonumber\\
& &\gamma_{0v}=-2\pi\alpha^{\prime}A^{\prime}_{0},
~~~\gamma^{00}=\frac{\gamma_{vv}}{\gamma_{00}\gamma_{vv}+\gamma_{0v}^{2}},~~~
\gamma_{vv}=\frac{\gamma_{00}}{\gamma_{00}\gamma_{vv}+\gamma_{0v}^{2}}.
\end{eqnarray}
One can obtain the following result for $A^{\prime}_{0}$ by setting
$E=H=0$ in~(\ref{4eq6}),
\begin{equation}
\label{5eq13}
A^{\prime2}_{0}=-\frac{C^{2}\gamma_{00}\gamma_{vv}}{\gamma_{ii}^{p}+(2\pi\alpha^{\prime})^{2}C^{2}}.
\end{equation}
Finally by combining~(\ref{5eq11}),~(\ref{5eq12}) and~(\ref{5eq13}),
we can arrive at
\begin{equation}
\label{5eq14} \sigma=\sqrt{(2\pi\alpha^{\prime})^{4}\tau_{\rm
eff}^{2}(\frac{L}{v_{+}})^{2p-4}+
(\frac{2\pi\alpha^{\prime}}{L^{2}})^{2}(J^{t})^{2}v_{+}^{4}},
\end{equation}
where we have used the relation
$J^{t}=(2\pi\alpha^{\prime})^{2}\tau_{\rm eff}C$. One can compare
this result with~(\ref{4eq18}). From~(\ref{4eq14}) we can see that
in the limit of $E\rightarrow0$, $v_{\ast}\rightarrow v_{+}$.
Then~(\ref{5eq14}) is in agreement of~(\ref{4eq18}), which also
provides a non-trivial check of consistency. Furthermore, in the
zero density limit $J^{t}\rightarrow0$, the conductivity is given by
\begin{equation}
\sigma=\tau_{\rm
eff}(2\pi\alpha^{\prime})^{2}\frac{L^{p-2}}{v_{+}^{p-2}},
\end{equation}
which agrees with the result obtained in~\cite{Pang:2009dc}.

When a non-trivial dilaton is considered, the conductivity can be
written as
\begin{equation}
\label{5eq16}
\tilde{\sigma}=\tilde{\mathcal{N}}e^{-\phi}\sqrt{-\gamma\gamma^{00}\gamma^{vv}}
\gamma^{ii}\big|_{v\rightarrow v_{+}},
\end{equation}
where
$\tilde{\mathcal{N}}=(2\pi\alpha^{\prime})^{2}\tilde{\tau}_{\rm
eff}$. Now the solution for $A_{0}^{\prime}$ is given by
\begin{equation}
A^{\prime2}_{0}=-\frac{C^{2}\gamma_{00}\gamma_{vv}}{e^{-2\phi}\gamma_{ii}^{p}+(2\pi\alpha^{\prime})^{2}C^{2}}.
\end{equation}
Then the conductivity reads
\begin{equation}
\label{5eq18}
\tilde{\sigma}=\sqrt{e^{-2\phi_{+}}(2\pi\alpha^{\prime})^{4}\tilde{\tau}_{\rm
eff}^{2}(\frac{L}{v_{+}})^{2p-4}+
(\frac{2\pi\alpha^{\prime}}{L^{2}})^{2}(J^{t})^{2}v_{+}^{4}},
\end{equation}
which agrees with~(\ref{4eq23}) once again in the limit of
$E\rightarrow0$. Notice that~(\ref{5eq18}) reduces to~(\ref{5eq14})
when the dilaton becomes trivial $\phi=\phi_{0}$.
\subsection{Charge diffusion constant}
Generally speaking, the charge diffusion constant can be read off
from the denominator of the $tt$-component of the longitudinal
correlator. However, since it is quite difficult to solve the
equations of motion for longitudinal fluctuations, even in the
hydrodynamic expansions, we cannot expect that the charge diffusion
constant can be obtained in that way. An alternative formula for
estimating the charge diffusion constant was derived
in~\cite{Kovtun:2003wp} from the membrane paradigm. Such a formula
was generalized to D$p$/D$q$ systems in~\cite{Mas:2008qs} by
considering the equation of motion for the gauge invariant
perturbations
$E_{\parallel}=q\mathcal{A}_{0}+\omega\mathcal{A}_{1}$. At zero
baryon density the corresponding equation of motion is
\begin{equation}
E_{\parallel}^{\prime\prime}+\partial_{r}\log[\frac{e^{-\phi}\sqrt{-\gamma}
\gamma^{ii}\gamma^{rr}}{\omega^{2}+q^{2}\frac{\gamma^{ii}}{\gamma^{00}}}]E_{\parallel}^{\prime}
-\frac{\omega^{2}\gamma^{00}+q^{2}\gamma^{ii}}{\gamma^{rr}}E_{\parallel}=0.
\end{equation}

Assuming a dispersion relation of the form $\omega=-iDq^{2}+\cdots$,
the natural hydrodynamic scaling is given by
$\omega\rightarrow\lambda^{2}\omega, q\rightarrow\lambda q$. Then
the near horizon expansion can be written as
\begin{equation}
E_{\parallel}(r)=F(r)^{-i\frac{\lambda^{2}\omega}{4\pi
T}}(E^{(0)}_{\parallel,\rm reg}+\lambda^{2}E^{(2)}_{\parallel,\rm
reg}+\cdots)=E^{(0)}_{\parallel}+\lambda^{2}E^{(2)}_{\parallel}+\cdots
\end{equation}
where $F(r)=(r-r_{H})F_{\rm reg}(r)$ and the subscript ``reg''
stands for the regular part of the corresponding function. The
solution at lowest order is given by
\begin{equation}
E^{(0)}_{\parallel}=1-iC_{0}\frac{q^{2}}{2\pi\omega
T}\int^{r}_{r_{H}}
\frac{dr}{e^{-\phi}\sqrt{-\gamma}\gamma^{00}\gamma^{rr}},~~~
C_{0}=-\frac{1}{2}e^{-\phi}\sqrt{-\gamma}\gamma^{00}\gamma^{rr}F^{\prime}\big|_{r\rightarrow
r_{H}}.
\end{equation}
Notice that the retarded correlation function
$G^{R}_{\parallel}\sim\lim_{r\rightarrow
r_{B}}E_{\parallel}^{\prime}(r)/E_{\parallel}(r_{B})$ where $r_{B}$
denotes the boundary. Then the dispersion relation comes from
requiring $E_{\parallel}(r_{B})=0$. Finally the charge diffusion
constant is determined as
\begin{equation}
\label{5eq22}
D_{0}=e^{-\phi}\sqrt{-\gamma\gamma^{00}\gamma^{rr}}\gamma^{ii}\big|_{r_{+}}
\int^{r_{B}}_{r_{+}}\frac{dr}{e^{-\phi}\sqrt{-\gamma}\gamma^{00}\gamma^{rr}},
\end{equation}
which reduces to the one in~\cite{Kovtun:2003wp} when $A_{0}=0$. The
charge diffusion constant for massive charge carriers at finite
baryon density is more complicated~\cite{Mas:2008qs}. However, for
massless charge carriers at finite density the charge diffusion
constant is still given by~(\ref{5eq22}).

Let us return to our concrete example. Since the integral
in~(\ref{5eq22}) involves the dilaton and we do not give its
explicit form, we consider the trivial dilaton case instead. After
substituting~(\ref{5eq12}) and~(\ref{5eq13}) into~(\ref{5eq22}), we
can obtain
\begin{equation}
\label{5eq23}
D_{0}=\frac{1}{(p-z)v_{+}^{z-2}}\sqrt{1+(\frac{2\pi\alpha^{\prime}}
{L^{p}})^{2}C^{2}v_{+}^{2p}}{}_{2}F_{1}[\frac{p-z}{2p},\frac{3}{2};
\frac{3}{2}-\frac{z}{2p};-(\frac{2\pi\alpha^{\prime}}{L^{p}})^{2}C^{2}v_{+}^{2p}].
\end{equation}
It can be seen that the above result reduces to the one obtained
in~\cite{Pang:2009dc} when the charge density parameter $C=0$.
Another point is that the charge diffusion constant becomes
divergent when $z=p$, which was also observed in~\cite{Pang:2009dc}
for zero density case. As a check of consistency, we can choose one
specific example $z=1, p=3$ and~(\ref{5eq23}) becomes
\begin{equation}
D_{0}=\frac{1}{2\pi
T}\sqrt{1+d^{2}}{}_{2}F_{1}[\frac{1}{3},\frac{3}{2};\frac{4}{3};-d^{2}],~~~
d=2\pi\alpha^{\prime}Cv_{+}^{3}/L^{6},
\end{equation}
which agrees with the result for D3/D7 in~\cite{Kim:2008bv}.

One may wonder if the Einstein relation still holds for our system.
The charge susceptibility is determined by
\begin{equation}
\chi=\frac{\partial n_{q}}{\partial\mu}\big|_{T},
\end{equation}
where $n_{q}=\delta S_{q}/\delta A^{\prime}_{0}$ is the charge
density and $\mu$ denotes the chemical potential. Then
\begin{equation}
\chi=(\int^{r_{B}}_{r_{H}}\frac{dA^{\prime}_{0}}{dn_{q}}dr)^{-1}.
\end{equation}
It has been observed in~\cite{Mas:2008qs} that for massless charge
carriers the charge susceptibility has a simple form
\begin{equation}
\label{5eq27}
\chi=\mathcal{N}(\int^{r_{B}}_{r_{H}}\frac{1}{e^{-\phi\sqrt{-
\gamma}\gamma^{00}\gamma^{rr}}}dr)^{-1}.
\end{equation}
Therefore by combining~(\ref{5eq22}) and~(\ref{5eq27}), we can
conclude that the Einstein relation $D_{0}=\sigma/\chi$ holds for
massless charge carriers.
\section{Remarks on massive charge carriers}
In this section we will give some remarks on the properties of
massive charge carriers. In general, the properties for massive
charge carriers exhibit qualitatively similar behavior as their
massless counterparts, so we will just present the main results in
brief.
\subsection{Thermodynamics}
For massive charge carriers the embedding profile of the probe
D$q$-brane is described by a non-trivial function $\theta(r)$. Then
the induced metric on the D$q$-brane is given by
\begin{equation}
ds^{2}_{D_{q}}=-r^{2z}f(r)dt^{2}+(\frac{1}{r^{2}f(r)}+\theta(r)^{\prime2})dr^{2}
+r^{2}d\vec{x}^{2}_{p}.
\end{equation}
The effective action takes the following form for the trivial
dilaton case
\begin{equation}
S_{q}=-\tau_{\rm eff}\int
drr^{p}\cos^{n}\theta\sqrt{r^{2z-2}+r^{2}f(r)\theta^{\prime2}-A^{\prime2}_{0}}.
\end{equation}
Similarly the solution for $A^{\prime}_{0}$ is
\begin{equation}
A^{\prime}_{0}=\sqrt{\frac{r^{2z-2}+r^{2}f(r)\theta^{\prime2}}
{1+\frac{r^{2p}\cos^{2n}\theta}{\tilde{d}^{2}}}},
\end{equation}
where $\tilde{d}=\rho/\tau_{\rm eff}$ and
$\rho=\delta\mathcal{L}_{q}/\delta A^{\prime}_{0}$. It has been
pointed out in~\cite{Karch:2007br} that the equation of motion for
the embedding profile $\theta(r)$ is not analytically solvable
except in the limit of zero temperature.

We can study the free energy in canonical ensemble by performing the
Legendre transformation to the action
\begin{equation}
\Omega=\tilde{d}\tau_{\rm
eff}\int^{\infty}_{r_{+}}dr\sqrt{r^{2z-2}f(r)+r^{2}f(r)\theta^{\prime2}}
\sqrt{1+\frac{r^{2p}\cos^{2n}\theta}{\tilde{d}^{2}}}.
\end{equation}
Following~\cite{Karch:2009eb}, to investigate the low temperature
behavior of the free energy we may perform a low-temperature
expansion around $r_{+}=0$,
\begin{equation}
\Omega=\Omega(r_{+})\big|_{r_{+}=0}+(\frac{\partial\Omega}{\partial
r_{+}})\big|_{r_{+}=0}r_{+}+\mathcal{O}(r_{+}^{2}).
\end{equation}
After carefully evaluating the behavior of $\Omega$ at the horizon
and at the boundary, we can obtain the expansion similar to that
in~\cite{Karch:2009eb}
\begin{equation}
\Omega=\Omega_{0}+\rho r_{+}+\mathcal{O}(r_{+}^{2}).
\end{equation}
One crucial point that is different from~\cite{Karch:2009eb} is that
the free energy contributes to the entropy even at zero temperature,
which is the constant term given in~(\ref{3eq15}). Therefore the
specific heat is dominated by the subleading term, which is similar
to the cases discussed in~\cite{Karch:2007br}. For the case of a
non-trivial dilaton the conclusion is similar.

\subsection{Conductivity and drag force}
By assuming the same configuration of the worldvolume gauge fields
and following the same steps presented in section 4, we can arrive
the following results for the DC conductivity and the DC Hall
conductivity for the case of a trivial dilaton
\begin{equation}
\sigma=\sqrt{(2\pi\alpha^{\prime})^{4}\tau_{\rm
eff}^{2}\cos^{2n}\theta_{\ast}(\frac{L}{v_{\ast}})^{2p-4}
+(\frac{2\pi\alpha^{\prime}}{L^{2}})^{2}(J^{t})^{2}v_{\ast}^{4}},
\end{equation}
\begin{eqnarray}
\sigma^{xx}&=&\frac{1}{1+(\frac{2\pi\alpha^{\prime}}{L^{2}})^{2}B^{2}v^{4}_{\ast}}
\sqrt{(2\pi\alpha^{\prime})^{4}\tau_{\rm
eff}^{2}\cos^{2n}\theta_{\ast}(\frac{L}{v_{\ast}})^{2p-4}(1+(\frac{2\pi\alpha^{\prime}}{L^{2}})^{2}
B^{2}v_{\ast}^{4})+(\frac{2\pi\alpha^{\prime}}{L^{2}})^{2}(J^{t})^{2}v_{\ast}^{4}},\nonumber\\
\sigma^{xy}&=&\frac{BJ^{t}v_{\ast}^{4}}{(\frac{L^{2}}{2\pi\alpha^{\prime}})^{2}+B^{2}v_{\ast}^{4}}.
\end{eqnarray}
Notice that the quantities $\xi$ and $a$ are still given
by~(\ref{4eq31}) and~(\ref{4eq33}), while the expression for $\eta$
receives some modifications
\begin{equation}
\eta=-g_{tt}g_{xx}^{p}\cos^{2n}\theta-(2\pi\alpha^{\prime})^{2}[g_{tt}C^{2}+g_{xx}(H^{2}+M^{2})].
\end{equation}
Here $v_{\ast}$ is still determined by $\xi(v_{\ast})=0$. For the
case of a non-trivial dilaton,
\begin{equation}
\tilde{\sigma}=\sqrt{(2\pi\alpha^{\prime})^{4}e^{-2\phi_{\ast}}\tilde{\tau}_{\rm
eff}^{2}\cos^{2n}\theta_{\ast}(\frac{L}{v_{\ast}})^{2p-4}
+(\frac{2\pi\alpha^{\prime}}{L^{2}})^{2}(J^{t})^{2}v_{\ast}^{4}},
\end{equation}
\begin{eqnarray}
\tilde{\sigma}^{xx}&=&\frac{1}{1+(\frac{2\pi\alpha^{\prime}}{L^{2}})^{2}B^{2}v^{4}_{\ast}}
\sqrt{(2\pi\alpha^{\prime})^{4}e^{-2\phi_{\ast}}\tilde{\tau}_{\rm
eff}^{2}\cos^{2n}\theta_{\ast}(\frac{L}{v_{\ast}})^{2p-4}(1+(\frac{2\pi\alpha^{\prime}}{L^{2}})^{2}
B^{2}v_{\ast}^{4})+(\frac{2\pi\alpha^{\prime}}{L^{2}})^{2}(J^{t})^{2}v_{\ast}^{4}},\nonumber\\
\sigma^{xy}&=&\frac{BJ^{t}v_{\ast}^{4}}{(\frac{L^{2}}{2\pi\alpha^{\prime}})^{2}+B^{2}v_{\ast}^{4}}.
\end{eqnarray}
In this case $\xi$ and $a$ still remain invariant, while
$\tilde{\eta}$ becomes
\begin{equation}
\tilde{\eta}=-g_{tt}g_{xx}^{p}e^{-2\phi}\cos^{2n}\theta-(2\pi\alpha^{\prime})^{2}[g_{tt}C^{2}+g_{xx}(H^{2}+M^{2})].
\end{equation}
It can be seen that the conductivity for massive charge carriers
exhibits qualitatively similar behavior as their massless
counterparts, up to a factor of $\cos^{2n}\theta_{\ast}$ in the
first term inside the square root. Furthermore, the DC Hall
conductivity remains invariant for all cases. Therefore the
resistivity in different limits also exhibits similar behavior as
the massless case.

From a ``microscopic'' point of view, the conductivity is still
given by~(\ref{5eq11}) for the case of trivial dilaton
or~(\ref{5eq16}) for the case of non-trivial dilaton, which gives
\begin{equation}
\sigma=\sqrt{(2\pi\alpha^{\prime})^{4}\tau_{\rm
eff}^{2}(\frac{L}{v_{+}})^{2p-4}\cos^{2n}\theta_{+}
+(\frac{2\pi\alpha^{\prime}}{L^{2}})^{2}(J^{t})^{2}v_{+}^{4}},
\end{equation}
\begin{equation}
\tilde{\sigma}=\sqrt{(2\pi\alpha^{\prime})^{4}e^{-2\phi_{+}}\tilde{\tau}_{\rm
eff}^{2}(\frac{L}{v_{+}})^{2p-4}\cos^{2n}\theta_{+}
+(\frac{2\pi\alpha^{\prime}}{L^{2}})^{2}(J^{t})^{2}v_{+}^{4}}.
\end{equation}
We can find agreement with the ``macroscopic'' results in the
$E\rightarrow0$ limit once again.

In the large mass limit, the relations between the conductivity
and/or Hall conductivity and the drag force were clarified
in~\cite{Karch:2007pd} and~\cite{O'Bannon:2007in}. As shown
in~\cite{O'Bannon:2007in}, in the large mass limit the flavor
excitations are expected to be described as a collection of
quasiparticles, obeying the following equation of motion
\begin{equation}
\frac{d\vec{p}}{dt}=\vec{E}+\vec{v}\times\vec{B}-\vec{F}_{\rm drag},
\end{equation}
where $v$ stands for the quasiparticle velocity. In the steady state
$d\vec{p}/dt=0$, we have
\begin{equation}
F_{\rm
drag}=\sqrt{E^{2}+v^{2}B^{2}+2\vec{E}\cdot(\vec{v}\times\vec{B})}.
\end{equation}
The thermal pair creation should be suppressed in the large mass
limit, hence
\begin{equation}
J^{x}=J^{t}v_{x},~~~J^{y}=J^{t}v_{y}.
\end{equation}
By setting $\chi(v_{\ast})=0$ we can obtain $v^{2}=|g_{tt}|/g_{xx}$,
where the first term has been suppressed in the large mass limit.
Setting $\xi=a=0$ leads to the following results
\begin{equation}
E^{2}=\frac{1}{(2\pi\alpha^{\prime})^{2}}g_{xx}^{2}v^{2}+v^{2}B^{2},~~~
v_{y}=-v^{2}\frac{B}{E}.
\end{equation}
Thus $2\vec{E}\cdot(\vec{v}\times\vec{B})=-2B^{2}v^{2}$, which
determines the drag force
\begin{equation}
F_{\rm drag}=\frac{1}{2\pi\alpha^{\prime}}g_{xx}(v_{\ast})v.
\end{equation}
One can see that this is precisely the formula for the drag force
in~\cite{Herzog:2006gh, Gubser:2006bz, Herzog:2006se}.

One can also obtain the conductivity tensor via the drag force. By
setting $\xi=0$ we can obtain
\begin{equation}
f(v_{\ast})=\frac{(2\pi\alpha^{\prime})^{2}E^{2}v^{2z+2}_{\ast}}
{1+(2\pi\alpha^{\prime})^{2}B^{2}v_{\ast}^{4}}.
\end{equation}
Then
\begin{equation}
v^{2}=-\frac{g_{tt}}{g_{xx}}=\frac{(2\pi\alpha^{\prime})^{2}E^{2}
v_{\ast}^{4}}{1+(2\pi\alpha^{\prime})^{2}B^{2}v_{\ast}^{4}}.
\end{equation}
The components of the speed are given by
\begin{eqnarray}
&
&v_{y}=-v^{2}\frac{B}{E}=-\frac{(2\pi\alpha^{\prime})^{2}EBv_{\ast}^{4}}
{1+(2\pi\alpha^{\prime})^{2}B^{2}v_{\ast}^{4}},\nonumber\\
&
&v_{x}=\sqrt{v^{2}-v_{y}^{2}}=\frac{2\pi\alpha^{\prime}Ev_{\ast}^{2}}
{1+(2\pi\alpha^{\prime})^{2}B^{2}v_{\ast}^{4}}.
\end{eqnarray}
Finally, by combining $J^{i}=J^{t}v_{i}$ and
$J^{i}=\sigma^{ij}E_{j}, i,j=x,y$, one can arrive at
\begin{equation}
\sigma^{xx}=\frac{\frac{2\pi\alpha^{\prime}}{L^{2}}J^{t}v_{\ast}^{2}}
{1+(\frac{2\pi\alpha^{\prime}}{L^{2}})^{2}B^{2}v_{\ast}^{4}},~~~
\sigma^{xy}=\frac{BJ^{t}v_{\ast}^{4}}{(\frac{L^{2}}{2\pi\alpha^{\prime}})^{2}+B^{2}v_{\ast}^{4}}.
\end{equation}
One can check that these results agree with previous ones in the
large mass limit. The relations between drag force and DC
conductivity were discussed in~\cite{Fadafan:2009kb} and here we
generalize it to the case of conductivity tensor.
\subsection{Charge diffusion constant}
The charge diffusion constant becomes more subtle for massive charge
carriers, as studied in~\cite{Mas:2008qs}. For completeness we just
list the result here, for details see~\cite{Mas:2008qs}.

The charge diffusion constant for massive charge carriers with
trivial dilaton is given by
\begin{equation}
D=\frac{D_{0}}{1+\frac{n_{q}}{2\pi\alpha^{\prime}\tau_{\rm
eff}}\int\frac{\Delta\tilde{\Psi}^{\prime}_{(0)}
+\Xi\tilde{\Psi}_{(0)}}{\sqrt{-\gamma}\gamma^{00}\gamma^{rr}}},
\end{equation}
where $D_{0}$ is given by~(\ref{5eq22}) and
\begin{equation}
\Delta=\gamma^{rr}\psi^{\prime}g_{\psi\psi},~~~\Xi=\frac{1}{2}(\gamma^{rr}\psi^{\prime2}
g_{\psi\psi,\psi}-n\gamma^{\theta\theta}\gamma_{\theta,\psi})
\end{equation}
with $\psi\equiv\sin\theta$. $\tilde{\Psi}_{(0)}$ is related to the
zeroth order expansion of the fluctuation of the embedding profile
by $\tilde{\Psi}_{(0)}=q\Psi_{(0)}$. For the case of a non-trivial
dilaton we just replace $\tau_{\rm eff}$ by $\tilde{\tau}_{\rm
eff}$. It can be easily seen that $\Delta=\Xi=0$  and $D=D_{0}$ for
massless charge carriers.

The charge susceptibility also receives modifications due to the
fact that $\psi$ also depends on $n_{q}$. For the case of
non-trivial dilaton, it is given by
\begin{equation}
\chi=(2\pi\alpha^{\prime})^{2}\tilde{\tau}_{\rm
eff}\big(\int^{r_{B}}_{r_{+}}\frac{dr}{e^{-\phi}\sqrt{-\gamma}\gamma^{00}\gamma^{rr}}
[1+n_{q}(\Delta\frac{\partial\psi^{\prime}}{\partial
n_{q}}+\Xi\frac{\partial\psi}{\partial n_{q}})]\big)^{-1}.
\end{equation}
For the case of trivial dilaton one can replace $\tilde{\tau}_{\rm
eff}$ by $\tau_{\rm eff}$ and $e^{-\phi}$ by $e^{-\phi_{0}}$. This
expression reduces to the one given in~(\ref{5eq27}) for massless
charge carriers where $\Delta=\Xi=0$. The Einstein relation for
massive charge carriers should be checked numerically and it was
verified to be true for D3/D7 system in~\cite{Mas:2008qs}.

\section{Summary and discussion}
We study several properties of holographic strange metals in this
note. The dual bulk spacetime is $p+2$-dimensional Lifshitz black
holes and the role of charge carriers is played by probe
D$q$-branes. The cases of trivial dilaton and non-trivial dilaton
are discussed respectively. We calculate the free energy density and
chemical potential analytically for massless charge carriers,
expressing the results in terms of hypergeometric functions. The
entropy density and heat capacity at low temperature are also
obtained. We find that the heat capacity behaves like a gas of free
bosons at $z=2$ and like a gas of fermions at $z=2p$. It may
indicate a new type of quantum liquid for other cases. The speed of
sound at low temperature is also evaluated.

We calculate the DC conductivity and DC Hall conductivity in
$p+2$-dimensions, generalizing the four-dimensional results obtained
in~\cite{Hartnoll:2009ns}. There are always two terms inside the
square root of DC conductivity, one of which can be interpreted as
describing thermal pair production and is suppressed in the large
mass limit. The other term is proportional to the charge density
$J^{t}$ and takes a universal form in all dimensions. Furthermore,
the DC Hall conductivity also takes a universal form, which is
independent of $p$. We discuss the resistivity in different limits
and find that at large charge density, the resistivity is linear
dependent on temperature when $z=2$, while for the case of trivial
dilaton, the linear dependence cannot be realized for $p>2$ with
$z>0$ at vanishing charge density. When a non-trivial dilaton is
introduced, such a linear dependence can still be realized by
adjusting the parameters even at $p=2$.

As a check of consistency, we apply the formulae for conductivity
in~\cite{Mas:2008jz} to our backgrounds. Such formulae were derived
via Kubo formula then they could be seen as ``microscopic'' results.
We find that the ``microscopic'' results are in agreement with the
``macroscopic'' results obtained in section 4. We also obtain an
analytic result for the charge diffusion constant, which agrees with
the result for D3/D7 in specific limit. The Einstein relation
explicitly holds for massless charge carriers. We also give some
remarks on the corresponding properties of massive charge carriers.
In particular, the conductivity tensor can be reproduced via the
drag force calculations in the large mass limit.

There are several interesting directions which are worth
investigating. For massive charge carriers the fluctuations of the
embedding profile couple to the longitudinal fluctuations of the
worldvolume gauge fields, which results in highly complicated
differential equations. Recently a framework for calculating
holographic Green's functions from general bilinear actions and
fields obeying coupled differential equations in the bulk was
proposed in~\cite{Kaminski:2009ce, Kaminski:2009dh}. Then it would
be interesting to classify the vector and scalar fluctuations and
calculate the Green's functions for our system by employing their
method. Another point is to study the fermionic properties of the
background. In particular, it would be desirable to obtain the
Green's functions for the fermions, following~\cite{Gubser:2009dt,
Faulkner:2009am, Faulkner:2010da}. We expect to study such
fascinating topics in the future.
\bigskip \goodbreak \centerline{\bf Acknowledgments}
\noindent This work was supported by the National Research
Foundation of Korea(NRF) grant funded by the Korea government(MEST)
through the Center for Quantum Spacetime(CQUeST) of Sogang
University with grant number 2005-0049409.



\end{document}